# Funnel plots for visualizing uncertainty in the research performance of institutions[1]

Giovanni Abramo (*corresponding author*)
  *Laboratory for Studies of Research and Technology Transfer*
  *at the Institute for System Analysis and Computer Science (IASI-CNR)*
  *National Research Council of Italy*
    ADDRESS: Istituto di Analisi dei Sistemi e Informatica, Consiglio Nazionale delle Ricerche
    Via dei Taurini 19, 00185 Roma - ITALY
    tel. +39 06 7716417, fax +39 06 7716461, giovanni.abramo@uniroma2.it

Ciriaco Andrea D'Angelo
  *University of Rome "Tor Vergata"*
    ADDRESS: Dipartimento di Ingegneria dell'Impresa, Università degli Studi di Roma "Tor Vergata"
    Via del Politecnico 1, 00133 Roma - ITALY
    Tel. and fax +39 06 72597362, dangelo@dii.uniroma2.it

Leonardo Grilli
  *University of Florence – Italy*
    ADDRESS: Dipartimento di Statistica, Informatica, Applicazioni "G. Parenti", Università degli Studi di Firenze
    Viale Morgagni, 59, 50134 Firenze - ITALY
    Tel. +39 055 2751552 - fax +39 055 4223560, grilli@disia.unifi.it

**Abstract**

Research performance values are not certain. Performance indexes should therefore be accompanied by uncertainty measures, to establish whether the performance of a unit is truly outstanding and not the result of random fluctuations. In this work we focus on the evaluation of research institutions on the basis of average individual performance, where uncertainty is inversely related to the number of research staff. We utilize the funnel plot, a tool originally developed in meta-analysis, to measure and visualize the uncertainty in the performance values of research institutions. As an illustrative example, we apply the funnel plot to represent the uncertainty in the assessed research performance for Italian universities active in biochemistry.

---

# 1. Introduction

Data values almost always entail some degree of uncertainty. For example, survey values are usually provided with confidence intervals showing the likely range of the population values. Even when data are not obtained from surveys, as is the case in the evaluation of research performance, there is uncertainty due to a number of factors. For research performance, this is largely related to the assumptions and limits of the measurement instrument, and for aggregate measures, to the different sizes of the research units under consideration. Accounting for such uncertainty is crucial, to establish whether the performance of a unit is truly outstanding and not the result of random fluctuations. Indeed, the Royal Statistical Society recommends that performance reporting should always include measures of uncertainty, although in practice this is not always done (Bird et al., 2005). Indications of uncertainty are definitely not provided for the most popular yearly international university "league tables". This is true whether the rankings are produced by 'non-bibliometricians', such as the Shanghai Jiao Tong University Ranking (SJTU, 2014), QS World University Rankings (QS, 2014) and Times Higher Education World University Rankings (THE, 2014), or whether they are produced by bibliometricians themselves, such as the Scimago Institutions Ranking (Scimago, 2015). In our previous studies (Abramo et al., 2011a), the current authors are like others in omitting the provision of the likely range of research performance values for Italian universities. The CWTS Leiden Rankings instead indicate stability intervals (Waltman et al., 2012). However, in the vast bibliometrics literature there are indeed relatively few works dealing with uncertainty in research performance measures. Colliander and Ahlgren (2011) distinguish between stability intervals and confidence intervals. To them, confidence intervals reflect uncertainty about a population parameter, whereas stability intervals reflect uncertainty about the indicators calculated for the dataset at hand. Schneider (2013) warns against the use of statistical significance tests (NHST) in research assessments. Instead, he advocates informed judgment, free of the 'NHST ritual', in decision-making processes. Bornmann et al. (2013) reformulated the 2011/2012 Leiden ranking by means of multilevel regression models, earlier introduced and applied by Bornmann et al., (2011) and Mutz and Daniel (2007). Williams and Bornmann (2014) propose guidelines for the consideration of percentile-rank classes of publications, when analyzed by citation impact. These authors, drawing on work by Cumming (2012), show how examination of the effect of sizes and confidence intervals can permit clearer understanding of citation impact differences.

In this work we introduce the funnel plot as a tool to visualize the uncertainty in the research performance values of institutions. The funnel plot was originally developed in meta-analysis (Egger et al., 1997), and to the best of our knowledge has not been applied to bibliometric rankings of research performance. A funnel plot shows the uncertainty in data values by adding confidence bands, indicating the range where the true research performance value is likely to lie. The visualization of uncertainty is useful in both analyzing the data and communicating the results. The visualizations help: i) signpost that there is some uncertainty about the true values, and that the data values are subject to random fluctuations; ii) identify truly outstanding units, namely units whose difference from the overall mean is statistically significant; and iii) highlight which datasets are more reliable for decision-making, to permit for example giving more weight to datasets with smaller confidence bands (such as those based on larger samples). Finally, we provide an example of funnel plot application, utilizing it to visualize the uncertainty in the evaluation of research performance of Italian universities active in Biochemistry.

In the next section we describe the factors that are likely to cause distortion and uncertainty in research performance values. In Section 3 we illustrate the general use of the funnel plot as a tool for visualizing uncertainty. In Section 4 we present the dataset and the research performance indicator used in the analysis. In Section 5 we show the results from applying the funnel plot to our selected field of observation. Section 6 offers the conclusions.

**2. Uncertainty in the assessment of research performance**

According to Bougnol and Dulá (2015) all assessment processes are critically affected by 'subjective' aspects (arising mainly from assumptions about the operationalization of the performance indicator), as well as by purely technical issues (related to data handling). The literature on university rankings for educational performance (Guarino et al., 2005) proposes probabilistic approaches to treat the uncertainty and assess the statistical significance of differences across universities. Turning to measures of research performance, the inherent assumptions and limits again prompt the adoption of probabilistic approaches rather than deterministic ones, in reporting and interpreting the assessment results. A very important consideration is that any policy, administrative or operational decisions based on performance assessments should consider the uncertainty levels embedded in the results.

Bibliometric measures of performance are based on countable research output/impact, and in some cases also on inputs. Although the outputs captured by bibliometric indicators have well known limitations,[2] bibliometricians agree that in certain fields they are reliable proxies of overall output. Similarly, notwithstanding their limitations[3], citations are considered a reliable proxy of the impact of scientific research, as long as there is a sufficient time lapse from the publication date to the observation of citations (Abramo et al., 2011b).

To begin, we first examine the output/impact-side factors that introduce uncertainty in bibliometric indicators. We expect the factors of uncertainty to affect the indicators randomly, meaning they will generate fluctuations without systematic effects in favor of or against particular groups of researchers.[4] We argue that the main sources of uncertainty are the following:

---

[2] The limitations of bibliometric indicators include the following: not all new knowledge produced can be codified into countable documents, and a part of it remaining intangible; not all documents are indexed in the bibliometric databases typically used in assessment exercises, such as WoS or Scopus.

[3] Measuring the impact of a publication through citations suffers from problems such as negative citations and 'network' citations (Glänzel, 2008).

[4] On the contrary, bias factors generate fluctuations with systematic effects. For example, in comparing professors belonging to different fields, gross aggregations may favor those belonging to a particular field. The analyst should make efforts to eliminate the effects of such bias factors, for example by making comparisons within homogeneous fields (Abramo et al., 2008) or by field-normalizing citations. Bias can be reduced, even if no method can accomplish the perfect fine-grained classification of scientist or the perfect overlapping of citation distributions when assessing a large number of fields (Zhang et al., 2014; Abramo et al., 2012a; Radicchi et al., 2008).

1) Variability in the intensity of production due to personal events (e.g. the researcher has occasional periods of additional teaching or managerial duties, family problems, etc.);
2) Variability in the intensity of production due to patterns in research projects (e.g. the researcher has variable access to funding, or is cyclically unproductive due to engaging in the early stages of long-term projects or frontier projects);
3) Variability in both the number of publications in the period under observation and citations, due to the typical editorial and indexing processes (considering that even within a single discipline there is high variability in the time length of the peer review process, and time elapsing from acceptance to ultimate online/printed publication, and from publication to indexing by bibliometric databases);
4) Variability in the number of citations due to other accidental facts (indeed, the time pattern of citations for a paper is a non-linear process highly influenced by events in the early stage, such as the paper being mentioned in a plenary session of an important conference, or the paper being published just a few weeks before a competing paper; also, the citations can be the accidental outcome of the search strategy employed to retrieve the literature on a given topic (Karlsson et al., 2015);
5) Variability in the number of citations due to errors in databases (e.g. failure to indicate issue and volume, incorrect author or institution names, missing links between text references and the list of cited literature).

As for the input-side factors, differences are likely in the production factors available to different individuals or working units, and the research performance measures should be normalized for these. Unfortunately, relevant data are not easily available, especially at the individual level. We thus see the very typical assumption of equality in the resources available to all individuals or units. A further assumption, unless specific data are available, is that the hours each individual devotes to research are more or less the same. Finally, as is also seen for their outputs, the value of researchers is not undifferentiated. This will be reflected in differing costs for their labor, varying among research staff both within and between the units. Lack of information on these types of input data gives further cause for uncertainty in performance assessments.

In spite of the numerous elements of randomness discussed above, traditional bibliometric assessments are largely based on deterministic models that fail to consider and visualize uncertainty. There is also the further issue that bibliometricians compare research institutions on the basis of the average performance indicators of their members. In this context, accounting for uncertainty involves the additional feature that the various uncertainty factors are aggregated at the institution level, so that the amount of uncertainty will be inversely related to the size. Indeed in all rankings, small organizations often fall at the extremes as a consequence of high variability, while on the contrary large ones are usually located in the centre, as a consequence of their low variability. This differential variability due to size can be effectively handled by the funnel plot methodology introduced in the next section.

## 3. The funnel plot method

The comparison of research institutions on the basis of the overall average of a bibliometric index computed for each researcher poses methodological problems resembling those encountered in the evaluation of hospitals. It therefore appears worthwhile to consider the methods developed in the healthcare framework. Indeed, it is common practice to assess hospitals on the basis of the averages of a performance indicator observed at the patient level, for example for performance in patient readmissions or survival. The uncertainty of the mean observed for a given hospital is related, among other factors, to the institution's volume, meaning the number of patients. As a consequence of their higher variability, small hospitals are often located at the extremes of the ranking. Therefore, comparisons among institutions of different volumes should account for the variability of the performance indicators.

A popular method for visualising uncertainty is the caterpillar plot, where the performance measures of the units are plotted in increasing order and endowed with confidence intervals (see Spiegelhalter, 2005, and the references therein). The lengths of the intervals summarize the uncertainty and a unit whose interval is above (below) zero is judged to have a performance significantly above (below) the overall mean. Even if a caterpillar plot is technically correct, it may not be effective in communicating the results because: i) it does not explicitly show the relationship between the level of uncertainty and the volume or size of the units, and ii) it leads the reader towards undue emphasis on the ranking of the units though the reliability of the ranking is not assessed (the exact position of a unit is often found to be highly uncertain). The funnel plot overcomes these limitations.

The funnel plot is a graphic device for visualizing the uncertainty in the performance measures of units as a function of their volume or size. It was originally developed in meta-analysis and later adapted to the comparison of institutions with different volumes, such as hospitals (Spiegelhalter, 2005; Ieva and Paganoni, 2014). The funnel plot has two elements: i) a scatter of institutional outcome (in our case, the institution's research performance) against size (number of researchers); ii) confidence bands around the overall mean to assess if the observed outcome is statistically significant at a given confidence level (e.g. 95%). As the size of the institution increases, the standard error of the outcome decreases, thus the confidence bands converge toward the overall mean of the outcome. Typically, the institution's outcome is the mean of the chosen performance indicator at the individual level, thus the standard error and the implied confidence bands are inversely proportional to the square root of the number of observations (size), yielding funnel-shaped bands. The funnel plots usually show most institutions as falling within the bands, meaning there is no evidence that their performance is anomalous (and also implying that rankings would be misleading). Attention should instead be focused on those institutions falling outside the bands, whose performance is likely to be truly outstanding and worthy of closer scrutiny.

It is apparent that the funnel plot, as typically applied for hospitals, could also be exploited for comparing research institutions on the basis of a bibliometric index of performance. In such a case the outcome is the mean of the index across the researchers of the institution, while the size is the number of researchers. In symbols, let $y_{ij}$ be a bibliometric index for researcher $i$ of institution $j$. It is assumed that the individual observations randomly vary around the institution's mean, namely

$$y_{ij} = \mu_j + e_{ij},$$

[1]

where $\mu_j$ is the underlying institution's mean and $e_{ij}$ are independent and identically distributed errors with mean 0 and standard deviation $\sigma$. The errors are intended to capture all the sources of variability in the individual outcome, including higher or lower performance with respect to the institution's mean and the uncertainty factors discussed in Section 2.

For the fixed-effects model [1], the observed mean performance across the $n_j$ researchers of institution $j$, denoted as $\bar{y}_j$, is an unbiased estimator of the underlying mean $\mu_j$ with standard error $\sigma/\sqrt{n_j}$. Under normality, the bands of the funnel plot are defined as:

$$\bar{y} \pm z_{\alpha/2} \frac{s}{\sqrt{n_j}}$$

[2]

where $\bar{y}$ is an estimate of the overall mean and $s$ is an estimate of the standard deviation $\sigma$, obtained after fitting model [1] by least squares. The symbol $z_{\alpha/2}$ denotes the value of the Normal distribution with probability $\alpha/2$ on the right tail. The funnel plot usually has two pairs of bands, i.e. internal bands with $z_{\alpha/2}=2$ corresponding to a confidence level of about 95%, and external bands with $z_{\alpha/2}=3$ corresponding to a confidence level of about 99.7%.

The confidence bands [2] are valid if the institutional means are approximately normally distributed, which could be justified by the Central Limit Theorem: indeed, the distribution of a mean approaches normality regardless of the distribution of the individual values. However in practice, to achieve a satisfactory approximation it is advisable to exclude those institutions with few researchers, and to apply a suitable transformation to the index. For example, in the application illustrated in section 5, the transformation is the natural logarithm of the shifted index, namely $y_{ij} = \log(x_{ij} + \delta)$, where $x_{ij}$ is the bibliometric index in the original scale. The positive constant $\delta$ operates a translation so that the transformation can also be applied to researchers whose index is equal to 0. The value of $\delta$ should be chosen in a way that the distribution of the institution's means is approximately symmetrical.

The analysis can be carried out with any statistical software (we use Stata 13). However, as discussed in the conclusions, the steps for constructing a funnel plot are so simple that they could also be implemented using a spreadsheet.

**4. The research performance indicator**

In the Italian university system all professors are classified in one and only one field, named scientific disciplinary sector (SDS), 370 in all. In order to illustrate the use of the funnel plot to measure and visualize uncertainty in universities' research performance measures, we consider the scientific production of all Italian professors classified in the SDS "Biochemistry" over the period 2008-2012.

We have extracted data on the faculty at each university in the SDS Biochemistry from the database on Italian university personnel, maintained by the Italian Ministry of Education, University and Research[5]. The scientific production of these 911 professors was extracted by the Italian Observatory of Public Research, a bibliometric database developed and maintained by the first two authors and derived under license from the Thomson Reuters WoS. Beginning from the raw data of the WoS, and applying a complex algorithm to reconcile the author's affiliation and disambiguation of the true identity of the authors, each publication (article, review and conference proceeding) is attributed (3% error - harmonic average of precision and recall) to the university professor or professors that produced it (D'Angelo et al., 2011).

---

[5] http://cercauniversita.cineca.it/php5/docenti/cerca.php, last accessed on July 9, 2015.

We measure the research performance by an indicator of productivity. We first measure the productivity of each professor, and then average the productivity values of the faculty at each university in the SDS Biochemistry. Most bibliometricians define productivity as the number of publications in the period under observation. Because publications have different values (impact), we prefer to adopt a more meaningful definition of productivity: the value of output per unit value of labor, all other production factors being equal. The latter recognizes that the publications embedding new knowledge have a different value or impact on scientific advancement, which can be approximated with citations. Because citation behavior varies by field, we standardize the citations for each publication with respect to the average of the distribution of citations for all the cited Italian publications indexed in the same year and the same WoS subject category.[6] Furthermore, research projects frequently involve a team, which is registered in the co-authorship of publications. In this case we account for the fractional contributions of scientists to outputs which is, in the case of the life sciences (e.g. Biochemistry), further signaled by the position of the authors in the list of authors. The unit value of labor is expressed by the individual salary of professors, information that is usually unavailable for reasons of privacy. In the Italian case we have resorted to a proxy. In the Italian university system, salaries are established at the national level and fixed by academic rank and seniority. Thus all professors of the same academic rank and seniority receive the same salary, regardless of the university that employs them. The information on individual salaries is unavailable but the salaries ranges for rank and seniority are published. Thus we have approximated the salary for each individual as the national average of their academic rank.

At the individual level, we measure the average yearly productivity, termed the fractional scientific strength (*FSS*), as follows:[7]

$$FSS = \frac{1}{w_R} \cdot \frac{1}{t} \sum_{i=1}^{N} \frac{c_i}{\bar{c}} f_i$$

[3]

where the symbols are defined as follows:

$w_R$ = average yearly salary of the professor[8]
$t$ = number of years of work by professor in period under observation
$N$ = number of publications by professor in period under observation
$c_i$ = citations received by publication $i$
$\bar{c}$ = average of distribution of citations received for all cited publications in same year and subject category of publication $i$
$f_i$ = fractional contribution of professor to publication $i$.

---

[6] Abramo et al. (2012a) demonstrated that the average of the distribution of citations received for all cited publications of the same year and subject category is the best-performing scaling factor.

[7] A more extensive theoretical dissertation on how to operationalize the measurement of productivity can be found in Abramo and D'Angelo (2014).

[8] We adopt the following salary normalization coefficients: 1 for assistant; 1.4 for associate; 2 for full professors (source DALIA - https://dalia.cineca.it/php4/inizio_access_cnvsu.php, last accessed on July 9, 2015).

The fractional contribution is measured giving different weights to each co-author according to their position in the list of authors and the character of the co-authorship (intra-mural or extra-mural) (Abramo et al., 2013). If the first and last authors belong to the same university, 40% of the citation is attributed to each of them, the remaining 20% is divided among all other authors. If the first two and last two authors belong to different universities, 30% of the citation is attributed to the first and last authors, 15% of the citation is attributed to the second and penultimate, the remaining 10% is divided among all others[9].

## 5. A case application of the funnel plot

In this section we apply the funnel plot to measure and visualize the uncertainty of productivity values for Italian universities active in Biochemistry. The performance measure used is the FSS index, as defined in [3]. For reliable measurement of research performance, any index should be calculated over a sufficiently long period (Abramo et al., 2012b), thus we excluded those professors with less than three years on faculty over the observed period. We also excluded those universities with very small Biochemistry faculties (less than five professors). The final dataset of the analysis then consists of 42 universities and 877 professors. The distribution of FSS for the professors under observation is very skewed (skewness 3.14), with a mean value of 0.25, median of 0.13 and standard deviation of 0.34. In 37 cases we register nil values of FSS. The maximum value of FSS (2.95) is observed for a full professor at Polytechnic University of Milan. The 99[th] percentile of professors shows a much lower FSS (1.61), while the 95[th] is further halved (0.80).

To obtain the research productivity of each university at the SDS level (in our case, Biochemistry), we compute the mean performance of the university faculty in that SDS. The basic descriptive statistics of average FSS for the 42 universities are: mean 0.23, median 0.20, standard deviation 0.11, min-max 0.05-0.52.

As noted in Section 3, the confidence bands are valid if the institution's means are approximately normally distributed. To achieve a satisfactory approximation the recommended practice would be to exclude not only the institutions with very few professors (in our case less than 5) but also to apply a log transformation, shifting the index by a constant $\delta$. The value of $\delta$ is selected so that the distribution of the transformed data is approximately symmetric: specifically, $\ln(\text{FSS}+ \delta)$ is a zero-skewness log transform (Box and Cox, 1964). In our application, the selected value is $\delta = 0.01678$. We then compute the 42 institutions' means of the transformed index $\bar{y}_j$. Checking the normality of the distribution by means of a normal quantile plot (Figure 1), we see there is no evident deviation from normality.[10]

---

[9] The weightings were assigned on the basis of advice from senior Italian professors in the life sciences. The values could be changed to suit different practices in other national contexts.

[10] The institution's means $\bar{y}_j$ have different standard deviations, which are inversely proportional to $\sqrt{n_j}$; therefore, in order to apply normality checks on variables with a common standard deviation, the means have been adjusted using the formula $\sqrt{n_j}(\bar{y}_j - \bar{y})$.

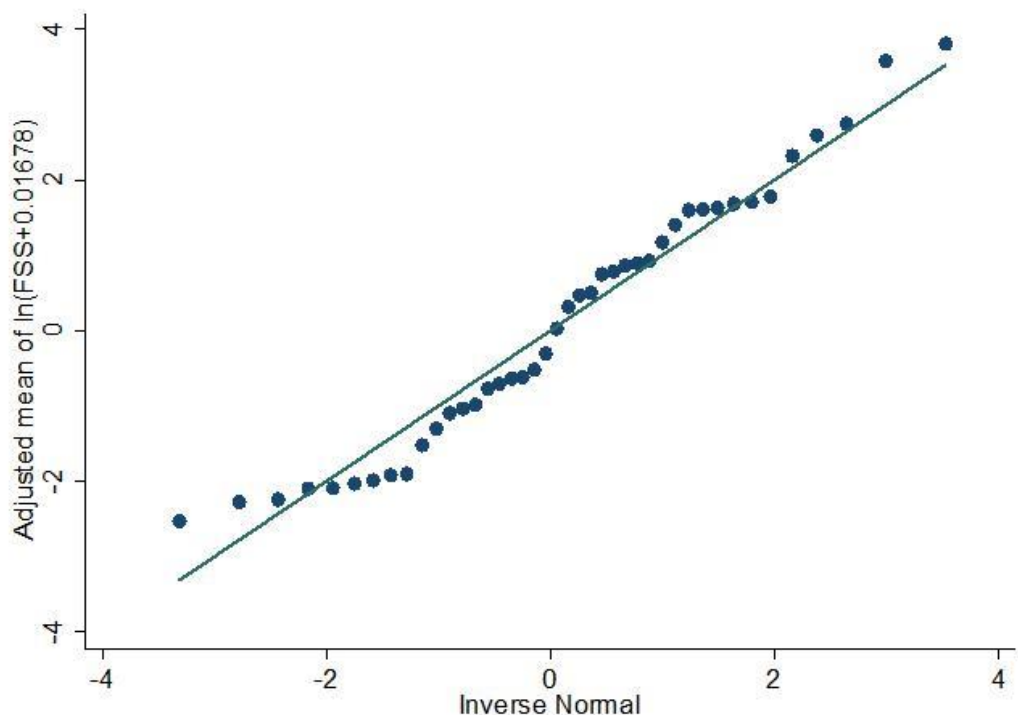

*Figure 1: Normal quantile plot of the distribution of adjusted means of the transformed FSS for 42 Italian universities with at least 5 professors in Biochemistry over the 2008-2012 period*

We can now graph the funnel plot (Figure 2) representing the distribution of the average performance of the universities $\bar{y}_j$ with the bands defined in [2] for $z_{\alpha/2}=2$ and $z_{\alpha/2}=3$. For performance in Biochemistry over the period 2008-2010, the plot shows that most universities lie within the bands defined for two standard deviations (two-SD bands) (33 out of 42, 79%) and all lie within the three-SD bands, with the exception of one university which is barely above the upper band. The plot thus points out some cases where performance can be confidently judged to be higher or lower than the average, but there are no instances of markedly outstanding performance. In the light of a funnel plot like the one in Figure 2, reporting a deterministic ranking of the universities has little meaning and may even be misleading.

Four universities are located above the upper two-SD band: the smallest one is the University of Insubria, with 11 professors and an average value of the transformed indicator equal to -0.93, the highest of the 42 assessed universities. Other universities with a performance above the upper two-SD band are Padua (47 professors), Turin (28 professors), and Bologna (41 professors), all with performance between -1.45 and -1.58. Padua is in fact the only university lying above the upper three-SD band. We note that the University of Salento is second in the ranking list with a performance of -1.241, but because of its small size (only 5 professors) it does not exceed the two-SD upper band.

Five universities are located below the lower two-SD band. University of Tuscia and Magna Grecia University are the smallest of this group, with five and seven professors respectively, and also with the worst performance among all 42 assessed universities (respectively -3.03 and -2.85). Among mid-sized universities, two are located below the two-SD lower band, namely Perugia (26 professors) and Pisa (25), with average performances between -2.4 and -2.5.

Among the large universities only Rome 'La Sapienza' (61 professors) is below the lower two-SD band. The performance of this university (-2.27) would earn 30th place in the ranking. The plot shows no universities below the lower three-SD band. Chieti 'D'Annunzio', with an average performance equal to -2.648, is third last in the ranking for the disciplinary sector, but its small size (10 professors) places it above the lower two-SD band.

Finally, we note that the funnel plot is designed to illustrate how the size of the institution affects the variability of the performance measure. However a different question is whether the size affects average performance, meaning whether larger units tend to have better or worse results. Fitting a regression line across the points in Figure 2 yields a slope close to zero (estimate -0.0013 with standard error 0.0123). Therefore the size of the institution is relevant only in terms of its relationship to the uncertainty of the observed performance indicator. The absence of returns to scale in this specific case study is in keeping with the literature on the subject (Abramo et al., 2012c; Avrikan, 2001; Bonaccorsi and Daraio, 2005).

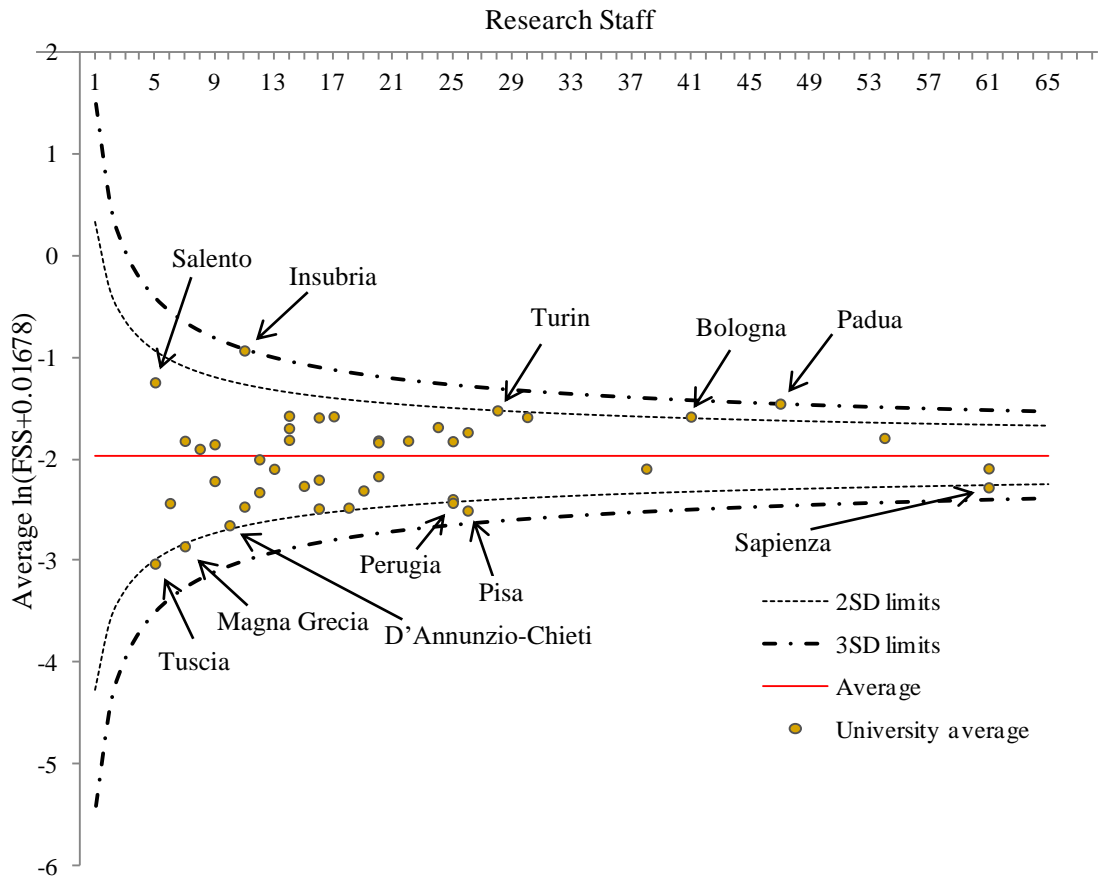

*Figure 2: Funnel plot of research productivity of 42 Italian universities with at least 5 professors in Biochemistry over the 2008-2012 period*

## 6. Conclusions

Uncertainty is embedded in any measurement process, and research performance measures offer no exception. Accounting for uncertainty is therefore obligatory, to establish whether the performance of an institution or research unit is truly outstanding or simply the result of random fluctuations. Bibliometricians generally apply deterministic methods to measure performance, neglecting the consideration of uncertainty. However, reporting the performance values of research units and institutions without uncertainty levels has little meaning and may even be misleading. This is especially so when the resulting ranking lists are used to inform policies and strategies, to guide the selective allocation of funding, or assist in other choices by stakeholders. The recent Leiden manifesto (Hicks et al., 2015) wisely recommends that practitioners 'avoid misplaced concreteness and false precision' in reporting performance values, and that 'if uncertainty and error can be quantified, for instance using error bars, this information should accompany published indicator values'.

In this work we have presented the funnel plot methodology as a tool to measure and visualize uncertainty, and have applied it to the case of performance measurement for Italian universities. The application demonstrates the key advantages of funnel plots for comparing performance, as summarized by Spiegelhalter (2005): 'there is no spurious ranking of institutions; the eye is naturally drawn to important points that lie outside the funnels; there is allowance for increased variability of the smaller units; the axes are easily interpretable, so additional data points can be added by hand; repeated observations over time can be plotted on the funnel and joined-up to show progress; easy to produce with standard spreadsheet programs'.

It is worth noting that the calculations for the funnel plot are very simple. Indeed, similarly to what has been done in evaluation of health institutions in England,[11] the procedure could be implemented in a downloadable spreadsheet, making it very widely available. The funnel plot is furthermore a general tool that can be used with any bibliometric indicator of performance. In our case we have used it with the FSS indicator, but it could also be used with the h-index, MNCS, number of publications, or other indicators.

From a statistical point of view, the funnel plot amounts to a series of tests of the null hypothesis that the mean performance of the institution under consideration is equal to the overall mean performance. These tests have a type I error rate defined by the bands, for example the two-SD bands correspond to a confidence level of about 95% and a type I error rate of about 5%. We therefore expect that for every 100 institutions not differing from the overall mean, the tests wrongly identify five of them as outlying. Clearly, using three-SD bands the type I error rate is very low (0.3%). However, like any testing procedure controlling for type I error, the funnel plot methodology does not control for type II error, namely the error of failing to detect an institution that truly differs from the overall mean as an outlier. Accordingly to the theory of statistical tests, the type II error rate is inversely related to sample size, meaning the number of professors at the institution. Consequently, the funnel plot methodology could have low power for the detection of outlying small-sized institutions. This feature is unavoidable, thus it would be desirable that future research investigate the behavior of the type II error rate in realistic scenarios, in order to give guidelines. The results of our example analysis show that, according to the two-SD bands, around 80% of Italian universities have a performance in biochemistry research that is not significantly different from the mean, and that there are no instances of markedly outstanding performance. These findings may be partly influenced by the low power of the methodology to detect outlying small-sized institutions, since the conduct of our analysis at the field level implies that most universities have few professors. In spite of this, we note that the findings are indeed consistent with the Italian context. In fact, unlike higher education systems such those of the U.S. and U.K., where fierce competition has led to the growth of 'outlier' world-class universities, in non-competitive (Italian and other) systems the variability of performance between universities is much lower than it is within them (Abramo et al., 2012d). In the near future we intend to extend the current analysis to all fields of research, with the aim of identifying those fields where outstanding performance occurs. We also intend to carry out the same analysis at the macro-disciplinary level (mathematics, physics, medicine, etc.) and at the overall university level.

---

[11] See for example Spiegelhalter blog: http://understandinguncertainty.org/fertility, last accessed on July 9, 2015.